\documentclass[%
 aip,
 amsmath,amssymb,
 reprint,%
]{revtex4-1}

\usepackage{graphicx}% Include figure files
\usepackage{dcolumn}% Align table columns on decimal point
\usepackage{bm}% bold math
\usepackage{soul}
\usepackage{xcolor}
\sethlcolor{yellow}

\soulregister\cite7
\soulregister\ref7
\soulregister\eqref7
\usepackage[utf8]{inputenc}
\usepackage[T1]{fontenc}
\usepackage{mathptmx}
\usepackage{etoolbox}

\makeatletter
\def\@email#1#2{%
 \endgroup
 \patchcmd{\titleblock@produce}
  {\frontmatter@RRAPformat}
  {\frontmatter@RRAPformat{\produce@RRAP{*#1\href{mailto:#2}{#2}}}\frontmatter@RRAPformat}
  {}{}
}%
\makeatother
\DeclareUnicodeCharacter{0301}{\'{}}
\begin{document}

\preprint{AIP/123-QED}

\title[]{Spin-wave phase modulation using magnetic domain walls in dipolarly coupled structures for non-volatile magnonic computation  }
\author{H. Mortada*}
 \affiliation{Fachbereich Physik and Landesforschungszentrum OPTIMAS, Rheinland-Pf\"alzische Technische Universit\"at Kaiserslautern-Landau, 67663 Kaiserslautern, Germany.}
\affiliation{%
Universit\'e Paris-Saclay, Centre de Nanosciences et de Nanotechnologies, CNRS, 91120, Palaiseau, France.
}%
 \email{hanadi.mortada@rptu.de}
\author{P. Pirro}
 \affiliation{Fachbereich Physik and Landesforschungszentrum OPTIMAS, Rheinland-Pf\"alzische Technische Universit\"at Kaiserslautern-Landau, 67663 Kaiserslautern, Germany.}
\author{A. Hamadeh}
\affiliation{%
Universit\'e Paris-Saclay, Centre de Nanosciences et de Nanotechnologies, CNRS, 91120, Palaiseau, France.
}%

\date{\today}

\begin{abstract}
A controllable phase shifter is a key component for spin-wave–based logic and information-processing devices. Here, we propose a domain-wall-position–controlled spin-wave phase shifter that exploits dipolar coupling between two closely spaced waveguides to enable continuous phase tuning over a range approaching  360$^\circ$ while keeping the spin-wave amplitude constant. Using micromagnetic simulations, we model a bias-free hybrid structure composed of a  nanoscale waveguide magnetostatically coupled to a half-ring-shaped structure both made from bismuth-doped yttrium iron garnet with strong perpendicular magnetic anisotropy. Displacing a domain wall in the half-ring modulates the dispersion relation in the adjacent straight waveguide due to the changed magnetostatic interaction, providing a compact and dynamically reconfigurable phase-shifting mechanism. This approach offers precise and non-volatile control over spin-wave propagation and is compatible with energy-efficient magnonic logic architectures. 
\end{abstract}

\maketitle

%Introduction

The field of magnonics has advanced rapidly from fundamental theoretical models toward the realization of functional on-chip devices, including magnonictransistors\cite{khitun2011non,chumak2014magnon}, logic gates\cite{ustinov2019nonlinear,schneider2008realization,vogt2014realization,klingler2014design}, waveguides\cite{sadovnikov2018magnon,chumak2010all,bailleul2003propagating,au2011excitation,yu2016approaching}, and phase shifters\cite{morozova2020magnonic,nikitin2014all,lake2022microscopic,louis2016bias,ustinov2021induced,zhang2019bias,zhu2014magnonic,au2012nanoscale} whose computational frameworks capitalize on encoding information not only in the amplitude but also in the phase of carrier signals. In this context, phase-encoded SW logic has attracted considerable attention because it enables logic inversion through the addition of a $\pi$ phase shift to the propagating signal. For practical phase-based magnonic logic, phase control should be achieved with minimal SW attenuation and broadband operational capability. Extensive research has focused on realizing tunable and reconfigurable magnonic phase shifters based on several phase-shifting mechanisms, including magnetoelectric coupling\cite{wang2018electric}, magnetostatic modulation via nanomagnets\cite{au2012nanoscale}, spin-current-induced torque control\cite{zhang2019bias}, geometric structuring\cite{dobrovolskiy2019spin}, and electric-field-induced strain or superconducting screening effects in hybrid structures such as ferrite–ferroelectric and ferrite–superconductor systems that enable modulation of the SW dispersion\cite{nikitin2014all,morozova2020magnonic}. In addition to extensive efforts directed toward the development of low-power SW phase shifters, experimental studies and micromagnetic simulations have demonstrated localized phase control of SWs through engineered magnetization textures\cite{toniato2022magnonic,au2012nanoscale,petrillo2024micromagnetic,wojewoda2023phase}.

Despite notable progress in the design and experimental realization of SW phase shifters based on applied magnetic fields or electric currents \cite{vogt2014realization,chumak2010all,rousseau2015realization,demidov2009control,kostylev2007resonant}, several fundamental challenges continue to limit their suitability for magnonic logic circuits compatible with nanoscale implementations.  First, many existing approaches rely on continuously applied magnetic fields or electric currents to achieve phase control, which inevitably leads to sustained energy consumption during operation. This requirement conflicts with the vision of low-power and energy-efficient magnonic logic architectures, where non-volatile operation is highly desirable. Second, the dependence on external bias fields and current lines poses significant obstacles to device miniaturization and hinders the seamless integration of magnonic components with CMOS-compatible integrated circuits. These considerations highlight the need for self-biased, nanoscale phase shifters that can exploit intrinsic magnetic textures or internal field inhomogeneities to control the SW phase without continuous power input. Magnetic domain walls (DWs) have therefore attracted considerable interest as reconfigurable elements for SW manipulation. 

Early theoretical and experimental studies demonstrated that DWs can act as localized phase shifters in magnonic interferometers \cite{han2019mutual,qin2021electric,zhao2020spin,yan2011all,hertel2004domain}, and that the accumulated SW phase depends sensitively on the DW internal spin texture and chirality \cite{buijnsters2016chirality}, particularly in the presence of interfacial Dzyaloshinskii--Moriya interaction (DMI), which gives rise to a geometric phase contribution. Moreover, SWs can exert spin-transfer torque (STT) on DWs via magnonic spin currents, enabling reciprocal SW--DW interactions \cite{fan2023coherent,han2019mutual,chang2018ferromagnetic,kim2012interaction}. However, most previously proposed DW-based SW phase shifters fundamentally rely on SWs traversing DWs located directly in the propagation path. This requirement introduces a third major limitation: DWs in the waveguide can lead to additional damping, reflection, pinning effects, and increased fabrication complexity, all of which are detrimental to robust phase control in practical circuits. In particular, forcing SWs to cross DWs restricts design flexibility and may compromise signal integrity in densely integrated magnonic logic networks.

In this work, we propose a magnonic phase-shifter concept that overcomes these limitations by decoupling the phase-control mechanism from the SW propagation channel. Our design exploits dipolar coupling between two closely spaced magnetic nanowaveguides, where a DW hosted in an adjacent half-ring waveguide modulates the local dispersion of SWs propagating in a straight waveguide. Micromagnetic simulations demonstrate that the resulting phase shift can be continuously tuned by the DW position. The proposed phase-shifting approach offers advantages over conventional methods based on localized magnetic-field inhomogeneities\cite{kostylev2007resonant}. First, it employs the static stray field of a DW rather than a current-carrying wire, enabling non-volatile, bias-free operation without continuous power consumption. Second, instead of relying on direct SW scattering—which leads to reflection and nonmonotonic transmission—the design utilizes dispersion engineering via dipolar coupling, allowing continuous and monotonic phase tuning with minimal backscattering and negligible amplitude loss. Third, the control element is geometrically decoupled from the propagation channel, with the DW located in an adjacent waveguide, thereby avoiding scattering, pinning, and defect-related losses. Finally, the architecture is more scalable and suitable for nanoscale integration, as it eliminates the need for current lines that suffer from Joule heating, electro-migration, and fabrication complexity, while enabling efficient DW control through spin currents or magnetic fields.

\begin{figure}
\centering
\includegraphics[width=\columnwidth]{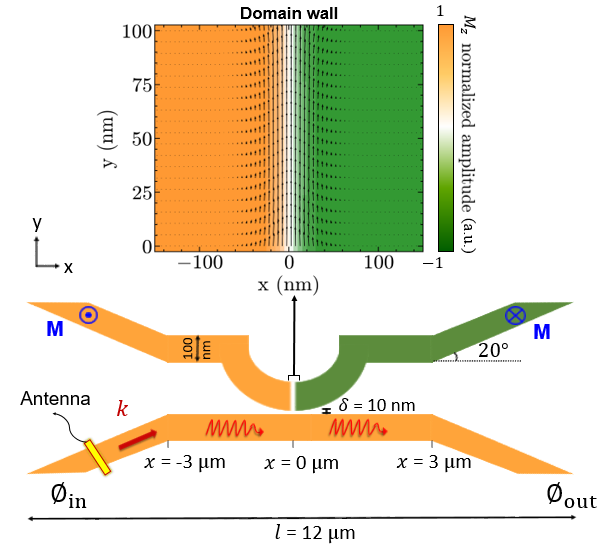}
\caption{\label{fig:1}Spin-wave phase shifter geometry. Schematic top view of the proposed Bi:YIG phase shifter consisting of a straight nanowaveguide dipolarly coupled to a half-ring waveguide separated by a 10 nm gap. SWs are excited at the left input port by a microwave antenna (yellow) at frequency $f = 5.9$ GHz with a fixed initial phase $\phi_{\text{in}}$. The half-ring hosts a magnetic DW whose position $x_{\mathrm{DW}}$ can be displaced along the ring. The color map shows the perpendicular magnetization component $M_z$ in the half-ring, illustrating the $180^\circ$ rotation of the out-of-plane magnetization across the DW.}
\end{figure}

% Micromagnetic simulations
 % Why Bi:YIG?
Among the most widely utilized low-damping magnetic materials in magnonic applications is Yttrium Iron Garnet (YIG). However, inducing perpendicular magnetization anisotropy (PMA) in YIG films presents a significant challenge, requiring substantial strain and generally being limited to film thicknesses below 50 nm.  One approach to overcoming this limitation involves doping YIG with elements exhibiting strong spin-orbit coupling, such as Bismuth (Bi), enabling fully out-of-plane magnetized films with thicknesses up to approximately 140 nm, making them well suited for nanostructured magnonic devices\cite{lin2020bi,chen202250}. Although Bi:YIG exhibits higher damping than pure YIG, it remains a low-damping magnetic insulator ($\alpha \approx 10^{-4}$) that supports long-range spin-wave propagation. This is confirmed in our study, where SWs propagate over distances exceeding 20 $\mu$m with exponential amplitude decay, well beyond the characteristic length scale of the present device ($\approx$10 $\mu$m). In addition, Bi:YIG enables energy-efficient DW manipulation via magnon spin currents. For instance, recent work by Fan et al.\cite{fan2023coherent} demonstrated the micrometer-scale manipulation of magnetic DWs within Bi:YIG, using coherently excited SWs as spin current carriers with amplitudes much smaller than metallic ferromagnetic systems, where DW motion typically requires high current densities on the order of 10$^{11}$–10$^{12}$ A/m$^2$, leading to significant Joule heating\cite{parkin2008magnetic,hayashi2008current}. In light of these advantages, Bi:YIG has been selected as the material of choice in this study for simulating the proposed device.

\begin{figure*}[t]
\centering
\includegraphics[width=\textwidth]{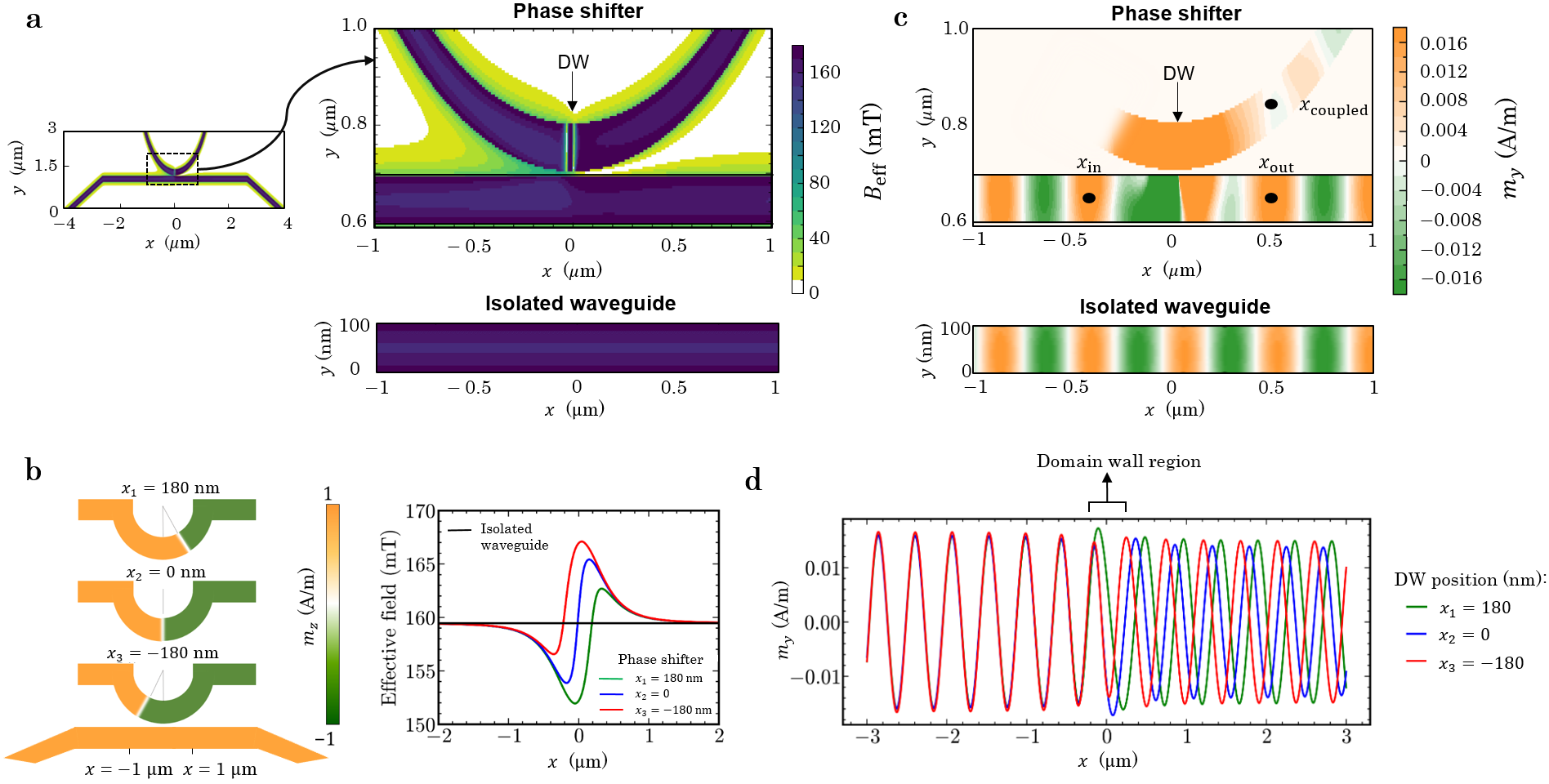}
\caption{\label{fig:2} \textbf{Domain–wall–induced effective-field modulation and its impact on spin-wave propagation.} \textbf{(a)} Spatial distribution of the effective magnetic field magnitude $B_{\text{eff}}$ in the straight waveguide dipolarly coupled to the half-ring hosting a DW at $x_{\mathrm{DW}} = 0$. For comparison, the effective-field distribution of an isolated straight waveguide (i.e., without the half-ring structure) is shown below. \textbf{(b)} One-dimensional line cuts of $B_{\text{eff}}$ extracted along the centerline of the straight waveguide ($y = 0.65~\mu$m, mid-width of the waveguide) for the isolated configuration (black) and for the coupled phase-shifter configuration at three representative DW positions: $x_{1}= +180$ nm (green), $x_{2}= 0$ nm (blue), and $x_{3}= -180$ nm (red).  \textbf{(c)} Snapshot of the dynamic magnetization component $m_y$ at time $t = 3$ ns for SWs excited at frequency $f = 5.9$ GHz. Top: coupled phase-shifter geometry with $x_{\mathrm{DW}} = 0$. Bottom: isolated straight waveguide. The antenna is located at $x_{\mathrm{antenna}} = -1~\mu$m. \textbf{(d)} SW profiles extracted along the propagation direction ($y = 0.65~\mu$m) for the three DW positions shown in (b), illustrating the DW-dependent phase evolution.}
\end{figure*}
% Device's parameters and dimensions:
The proposed phase shifter comprises a bent, nanoscale, self-biased SW waveguide magnetostatically coupled to a half-ring arranged in close proximity, as illustrated in Figure 1. Each of the two waveguides has a thickness "\(h\)" of 50 nm and a width "\(w\)" of 100 nm with the ring having a radius of 1.5 $\mu$m. The numerical modeling of this structure was conducted using the MuMax$^3$ micromagnetic simulation software package \cite{vansteenkiste2014design}, employing material parameters corresponding to Bi:YIG films: uniaxial anisotropy constant \(K_u = 12.5 \, \text{ kJ/m}^3\), exchange stiffness \(A_{\text{ex}} = 4.2 \, \text{ pJ/m}\), Gilbert damping coefficient \(\alpha = 1.3 \times 10^{-4}\), and saturation magnetization \(M_s = 100 \, \text{ kA/m}\) \cite{fan2023coherent}. No external bias magnetic field (\(\textbf{\textit{B}}_{\text{ext}}\)) was applied. To suppress back-scattering and the formation of standing waves, the Gilbert damping was exponentially increased to 0.5 at the waveguides ends. The exchange length is calculated as $\lambda_{\text{ex}} = \sqrt{\frac{2A_{\text{ex}}}{\mu_0 M_s^2}}$, yielding a value of 25.85 nm. All simulation cell dimensions were configured as $5 \, \text{nm} \times 5 \, \text{nm} \times 50  \, \text{ nm}$. A uniform out-of-plane magnetization profile along the thickness was assumed, which is an appropriate approximation for thin-film systems. The half-ring embeds a Bloch-type DW such that the magnetization direction rotates by 180$^\circ$ in the (yz)-plane; the upper inset shows the out-of-plane magnetization component profile $M_z(x,y)$ with the in-plane magnetization arrows indicating the Bloch rotation within the (y-z) plane. The localized rotation of the magnetization within the (yz)-plane minimizes stray-field energy, producing a stable Bloch DW configuration. The non-volatility of this design arises from the intrinsic stability of the DW in the half-ring geometry, where geometric confinement and the anisotropy landscape provide energy minima that pin the DW in well-defined positions.

The dipolar coupling between the lower waveguide and the half-ring structure is mediated across a narrow separation gap "\(\delta\)" of 10 nm, facilitating efficient coupling between the two conduits while maintaining geometrical confinement that supports single-mode FVSW propagation in the lower straight guide outside the coupling region. The selected gap represents a balance between maintaining strong dipolar coupling and minimizing the DW displacement needed to achieve phase inversion (phase shift of $180^\circ$); further details are discussed in the Supplementary Material. The spin-wave source (a localized field source) is a planar strip line with a width of 10 nm, positioned at the input port of the waveguide, as depicted in Figure 1 (shown in yellow). SW propagation is initiated by applying a sinusoidal magnetic field in time, \(b_y = b_0 \sin(2\pi ft)\), with an amplitude of \(b_0 = 1 \, \text{mT}\) and a selected microwave frequency \(f\) of 5.9 GHz. The antenna excites forward-volume SWs (FVSW), which subsequently traverse the coupling region and emerge from the output port with a modified phase. 

% Results
In the proposed design, the DW in the half-ring creates a localized perturbation of the magnetic environment, which is transferred non-locally to the straight waveguide by dipolar coupling, as illustrated in Figure 2. As a result, the effective field (i.e. the total local magnetic field acting on the magnetization, given by the sum of all field contributions included in the Landau–Lifshitz–Gilbert equation, namely the external (applied) field, exchange field, demagnetizing field, anisotropy field, and dipolar field contributions) experienced by propagating FVSWs becomes spatially inhomogeneous. Figure 2a illustrates the spatial distribution of the effective magnetic field magnitude $B_{\text{eff}}$ in the phase-shifter geometry, focusing on the region where the straight waveguide is dipolarly coupled to the half-ring waveguide hosting a DW. For comparison, the effective field profile of an isolated straight waveguide is also shown. In the isolated configuration, the effective field within the straight waveguide is nearly homogeneous, with a magnitude of approximately 159.3 mT, and exhibits no discernible variation along the SW propagation direction x. This uniform field reflects the single-domain magnetization state and results in a constant local dispersion relation for FVSWs. In contrast, when the half-ring waveguide is introduced, a pronounced spatial modulation of the effective field emerges in the straight waveguide. As shown in Figure 2a, the effective field in the coupled system varies from approximately 153 mT < $B_{\text{eff}}$ < 167 mT, corresponding to a peak-to-peak modulation of about $\Delta B_{\text{eff}}$ = 14 mT. This modulation is strongly localized beneath the DW region of the half-ring and extends over a lateral distance of approximately $\Delta x$ = 1 $\mu$m, defining the active phase-shifting section. Outside this region, the effective field rapidly recovers its uniform value, indicating that the perturbation does not affect the remainder of the waveguide.

The longitudinal effective field profile extracted along the centerline of the straight waveguide is presented in Figure 2b in an isolated case and in the phase shifter with three different DW positions in the half-ring, as indicated in the inset on the left. For the isolated waveguide (black curve), $B_{\text{eff}}$ remains constant at nearly 159.3 mT over the entire simulated length. In the phase-shifter configuration, the presence of the DW generates a strongly localized stray-field perturbation centered around the coupling region ($x$ = 0). The amplitude and spatial profile of this perturbation depend sensitively on the DW position. For a DW located at $x_{\text{DW}} = -180$ nm (red curve), the effective field exhibits a pronounced enhancement, reaching peak values of approximately 166 mT, followed by a gradual relaxation toward the reference field. Conversely, when the DW is displaced to $x_{\text{DW}} =+180$ nm (green curve), the field perturbation is inverted, producing a significant local reduction of the effective field down to approximately 152 mT. At the symmetric configuration $x_{\text{DW}}$ = 0 (blue curve), the effective field profile displays an intermediate behavior, with both a local minimum and maximum of reduced magnitude compared to the displaced cases. In all configurations, the field perturbation extends over a spatial range of several hundred nanometers, comparable to the characteristic wavelength of the propagating SWs. These results quantitatively demonstrate that DW displacement continuously reshapes the internal magnetic field landscape in the straight waveguide. Since the SW dispersion relation depends explicitly on the local effective field, the observed field modulation directly translates into a DW-position-dependent variation of the local wavenumber and, consequently, the accumulated SW phase. Thus, the tunable phase shift arises from controlled DW-induced stray-field engineering within the dipolar coupling region. 

The impact of the coupling and internal effective field modulation on SW dynamics is illustrated in Figure 2c, which compares the dynamic magnetization component $m_y (x,y,z)$ for the phase-shifter geometry and the isolated straight waveguide, measured at a fixed time instant at which the SW has reached the output port. In the isolated case, SWs excited by the antenna propagate with a constant wavelength, consistent with the uniform effective field and constant wavevector. In contrast, when the straight waveguide is coupled to the half-ring, a clear modification of the SW wavelength is observed within the coupled region. Specifically, the SW fringes become locally compressed and expanded as the wave traverses the region of reduced and enhanced effective field, respectively. Outside this region, the wavelength recovers its original value, confirming that the dispersion modification is spatially confined. The amplitude of the dynamic magnetization remains comparable to that of the isolated waveguide, indicating that the phase-shifting mechanism does not introduce significant attenuation or backscattering. To assess the energy efficiency of the proposed device and quantify transmission losses within the dipolarly coupled region, the SW power was evaluated at three specific locations: at the input port (\(x_{in}\)), at the output port (\(x_{out}\)), and at the corresponding output position along the half-ring waveguide (\(x_{coupled}\)), as indicated in Figure 2c. The SW power was computed from the time-averaged dynamic magnetization amplitude at the excitation frequency. Specifically, after performing a Fourier transform of the dynamic magnetization component $m_y(x,t)$, the power density at frequency $f$ was defined as $P(x) \propto |m_y(x,f)|^2$, averaged over the cross-sectional area of the waveguide and over several steady-state oscillation periods. This quantity is proportional to the local SW energy density in the linear regime, provides a consistent measure of transmitted SW intensity, and enables a direct comparison between the transmitted power in the main waveguide and the portion of energy transiently coupled into the half-ring. The results indicate that less than 10\% of the SW energy is dissipated within the half-ring, demonstrating that the device operates with high transmission efficiency and minimal energy loss through the dipolar interaction region. 

As the DW position changes along the half-ring, the spatial profile and magnitude of the dipolar coupling region vary, resulting in a controllable and tunable phase delay of the SW. Our simulations demonstrate that, by adjusting the DW position, the SW phase shift can be continuously tuned over a wide range, while maintaining a device geometry that is compatible with nanoscale implementations. Figure 2d shows the SW profiles extracted along the propagation direction for three different DW positions, $x_{\text{DW}}$ = $-180$ nm, 0 nm, and +180 nm. While the incoming SWs remain phase-aligned before entering the phase-shifting region, a clear relative phase offset develops after the waves exit the DW region. The accumulated phase shift depends systematically on the DW position, demonstrating that shifting the DW effectively changes the spatial extent over which the effective field, and thus the local wavevector, is modified. This behavior directly confirms that the phase shift arises from controlled dispersion engineering rather than from changes in propagation length or boundary reflections.
\begin{figure}
\centering
\includegraphics[width=0.7\columnwidth]{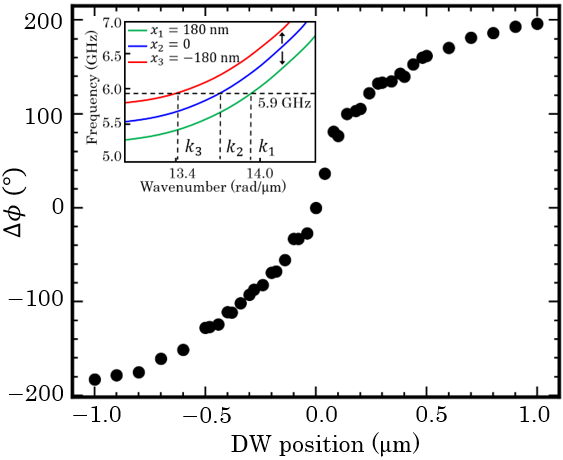}
\caption{\label{fig:3} \textbf{Quantitative extraction of the domain-wall-controlled spin-wave phase shift.} Extracted phase difference $\Delta \phi = \phi_{\text{in}} - \phi_{\text{out}}$ as a function of DW position $x_{\text{DW}}$, obtained from the phase of the FFT component at $f = 5.9$ GHz. The inset shows a demonstration of the effective field variation on the dispersion relation in the coupling region shown for three representative DW positions: $x_{1} = +180$ nm (green), $x_{2} = 0$ nm (blue), and $x_{3} = -180$ nm (red).}
\end{figure}

The phase difference between the input and output signals was determined via a frequency-domain methodology grounded in Fourier analysis. In dipolarly coupled waveguides, the SW wavenumber at a fixed frequency is highly sensitive to the local magnetization alignment, with regions dominated by parallel alignment supporting larger wavenumbers and consequently increased phase accumulation, whereas antiparallel alignment reduces the wavenumber and the accumulated phase\cite{mortada2025nonreciprocal}. When the DW is displaced to $x_1=+180$ nm, the half-ring and straight waveguide magnetizations are in a parallel configuration, yielding a reduced effective field in the coupling region and a corresponding negative shift in the dispersion relation, as shown in the inset of Figure 3. In contrast, when the DW is located at $x_2=0$ nm, it resides at the center of the half-ring with balanced parallel and antiparallel magnetization states, and for $x_3=-180$ nm, the magnetizations adopt an antiparallel configuration, increasing the effective field and producing a positive shift in the dispersion. These local modifications in the dispersion induce changes in the SW wavenumber at the excitation frequency, resulting in position-dependent phase shifts. Consequently, shifting the DW effectively controls the spatial extent over which the dispersion is modified, resulting in a tunable phase delay without altering the excitation frequency or applying an external bias field.  The phase difference is defined as $\Delta \phi = \phi_{\text{in}}-\phi_{\text{out}}$. We define the phase shift relative to a reference configuration corresponding to a centered DW ($x_{\text{DW}}=0$). In this symmetric state, the straight waveguide experiences equal spatial contributions from parallel and antiparallel dipolar coupling, resulting in a balanced modification of the local dispersion and a minimal additional phase accumulation. Figure 3 summarizes the extracted spin-wave phase difference, $\Delta \phi$, as a function of the DW position at the excitation frequency of 5.9 GHz. The phase response exhibits a continuous and strongly nonlinear dependence on the DW displacement $x_{\text{DW}}$, varying smoothly over a total range of approximately 360$^\circ$ across the investigated interval from $-1$ $\mu$m to +1 $\mu$m.  For negative DW displacements ($x_{\text{DW}}<0$), the accumulated phase shift is negative and reaches values close to $-180^\circ$ for large leftward offsets. As the DW approaches the central position, the phase decreases rapidly and crosses zero at  $x_{\text{DW}}=0$, by definition. Further displacement of the DW to the right ($x_{\text{DW}}>0$) results in increasingly positive phase shifts, approaching $+180^\circ$ at the largest positive offsets. The steep phase gradient near the central DW position indicates enhanced sensitivity in this regime, while the gradual saturation at large |$x_{\text{DW}}$| reflects the finite spatial extent of the dipolar coupling region. The asymmetric yet continuous dependence of  $\Delta \phi$ on $x_{\text{DW}}$ reflects the intrinsic asymmetry of the dipolar-field distribution in the bent geometry and confirms that the device operates as a dispersion-engineered, DW-position–controlled spin-wave phase shifter. 

Evidently, precise and stable positioning of DWs is critical in the proposed device, as the resulting phase shift is sensitive to DW location, necessitating fine control. While the DW position is predefined in the micromagnetic simulations through energy relaxation prior to SW excitation, previous experimental studies have demonstrated reliable DW stabilization and controllable displacement using magnetic fields, 
current-driven effects, and magnonic spin currents~\cite{fan2023coherent}. In particular, 
nanometer-scale DW repositioning has been experimentally achieved in systems with engineered pinning landscapes, enabling controlled transitions between stable pinning 
sites~\cite{parkin2008magnetic,hayashi2008current,wang2026coherent,tetienne2014nanoscale}. Consistent with these observations, additional micromagnetic simulations were performed to investigate magnon-driven DW dynamics in the proposed geometry. Within this geometry featuring strong PMA, the DW remains stably confined within a potential well and can be displaced with nanometer-scale precision by tuning the magnitude and duration of 
SW excitation, thereby enabling fine phase control while preserving non-volatility. To further assess DW stability, we systematically examined the micromagnetic energy landscape $E_\mathrm{tot}(x_\mathrm{DW})$ as a function of DW position, comparing a disorder-free waveguide with a realistic polycrystalline 
configuration. In the absence of disorder, the energy landscape is essentially flat, providing no intrinsic 
stabilization against thermal fluctuations at room temperature ($k_BT \approx 4.1~\mathrm{aJ}$). In contrast, the introduction of polycrystalline 
disorder — modeled as 150~nm grains with $\pm$10\% random variation in the uniaxial anisotropy constant $K_u$ — generates a strongly modulated pinning landscape with local energy barriers, ensuring robust thermal stability at each pinning site and suppressing 
spontaneous depinning. These results support the 
feasibility of achieving fine and reproducible phase control through precise DW manipulation. Further details regarding the experimental considerations, simulation 
parameters, and quantitative analysis of both DW motion and DW stability are provided in the Supplementary Material.

% Conclusion
In this work, we have presented a reconfigurable SW phase-shifting device that utilizes the controlled displacement of magnetic domain walls to achieve tunable phase modulation of propagating spin waves (SWs). The proposed architecture consists of a nanoscale, self-biased bent SW waveguide dipolarly coupled to a half-ring-shaped waveguide. Micromagnetic simulations reveal that isotropic forward-volume-like SWs propagating within this hybrid configuration accumulate a phase that varies systematically with the DW position. In contrast, the SW amplitude is almost independent of the DW position.  A pronounced phase shift is observed as the SW traverses the coupled region containing the DW, a phenomenon absent in otherwise identical configurations lacking the DW. The operational principle of the device relies on the positional tunability of the DW, which serves as a dynamic control parameter for reconfiguring the phase-shifting behavior. This mechanism enables precise, non-volatile, and reversible phase modulation without the need for a permanent external bias field, thereby offering a robust platform for energy-efficient and compact magnonic functionality.

\subsection*{SUPPLEMENTARY MATERIAL} See supplementary material for details of separation gap influence on the spin-wave phase shift and the fully micromagnetic simulations of domain wall stability in the proposed device and the spin wave-domain wall interaction in a Bi:YIG waveguide inducing domain wall motion.
\subsection*{ACKNOWLEDGEMENTS} This research has been supported by the European Union via the European Research Council (ERC) within the Starting Grant No. 101042439 "CoSpiN" and the Deutsche Forschungsgemeinschaft (DFG, German Research Foundation) - TRR 173 -268565370 (project B01).
\subsection*{DATA AVAILABILITY} 
The data that support the findings of this study are available from the corresponding author upon reasonable request.
\subsection*{CONFLICTS OF INTEREST} The authors declare no conflict of interest.

\section*{References}
\bibliography{references}

@article{khitun2011non,title={Non-volatile magnonic logic circuits engineering},author={Khitun, Alexander and Wang, Kang L},journal={Journal of Applied Physics},volume={110},number={3},year={2011},publisher={AIP Publishing}}

@article{ustinov2019nonlinear,title={Nonlinear spin-wave logic gates},author={Ustinov, Alexey B and L{\"a}hderanta, Erkki and Inoue, Mitsuteru and Kalinikos, Boris A},journal={IEEE Magnetics Letters},volume={10},pages={1--4},year={2019},publisher={IEEE}}

@article{chumak2010all,title={All-linear time reversal by a dynamic artificial crystal},author={Chumak, Andrii V and Tiberkevich, Vasil S and Karenowska, Alexy D and Serga, Alexander A and Gregg, John F and Slavin, Andrei N and Hillebrands, Burkard},journal={Nature communications},volume={1},number={1},pages={141},year={2010},publisher={Nature Publishing Group UK London}}

@article{sadovnikov2018magnon,title={Magnon straintronics: Reconfigurable spin-wave routing in strain-controlled bilateral magnetic stripes},author={Sadovnikov, AV and Grachev, AA and Sheshukova, SE and Sharaevskii, Yu P and Serdobintsev, AA and Mitin, DM and Nikitov, SA},journal={Physical review letters},volume={120},number={25},pages={257203},year={2018},publisher={APS}}

@article{chumak2014magnon,title={Magnon transistor for all-magnon data processing},author={Chumak, Andrii V and Serga, Alexander A and Hillebrands, Burkard},journal={Nature communications},volume={5},number={1},pages={4700},year={2014},publisher={Nature Publishing Group UK London}}

@article{schneider2008realization,title={Realization of spin-wave logic gates},author={Schneider, Thomas and Serga, Alexander A and Leven, Britta and Hillebrands, Burkard and Stamps, Robert L and Kostylev, Mikhail P},journal={Applied Physics Letters},volume={92},number={2},year={2008},publisher={AIP Publishing}}

@article{vogt2014realization,title={Realization of a spin-wave multiplexer},author={Vogt, Katrin and Fradin, Frank Y and Pearson, John E and Sebastian, Thomas and Bader, Samuel D and Hillebrands, Burkard and Hoffmann, Axel and Schultheiss, Helmut},journal={Nature communications},volume={5},number={1},pages={3727},year={2014},publisher={Nature Publishing Group UK London}}

@article{klingler2014design,title={Design of a spin-wave majority gate employing mode selection},author={Klingler, Stefan and Pirro, Philipp and Br{\"a}cher, Thomas and Leven, Britta and Hillebrands, Burkard and Chumak, Andrii V},journal={Applied Physics Letters},volume={105},number={15},year={2014},publisher={AIP Publishing}}

@article{bailleul2003propagating,title={Propagating spin wave spectroscopy in a permalloy film: A quantitative analysis},author={Bailleul, Matthieu and Olligs, Dominik and Fermon, Claude},journal={Applied Physics Letters},volume={83},number={5},pages={972--974},year={2003},publisher={American Institute of Physics}}

@article{au2011excitation,title={Excitation of propagating spin waves with global uniform microwave fields},author={Au, Y and Davison, T and Ahmad, E and Keatley, Paul Steven and Hicken, RJ and Kruglyak, VV},journal={Applied Physics Letters},volume={98},number={12},year={2011},publisher={AIP Publishing}}

@article{yu2016approaching,title={Approaching soft X-ray wavelengths in nanomagnet-based microwave technology},author={Yu, Haiming and d’Allivy Kelly, O and Cros, V and Bernard, R and Bortolotti, P and Anane, A and Brandl, F and Heimbach, F and Grundler, D},journal={Nature communications},volume={7},number={1},pages={11255},year={2016},publisher={Nature Publishing Group UK London}}

@article{zhang2019bias,title={Bias-free reconfigurable magnonic phase shifter based on a spin-current controlled ferromagnetic resonator},author={Zhang, Zikang and Liu, Shuang and Wen, Tianlong and Zhang, Dainan and Jin, Lichuan and Liao, Yulong and Tang, Xiaoli and Zhong, Zhiyong},journal={Journal of Physics D: Applied Physics},volume={53},number={10},pages={105002},year={2019},publisher={IOP Publishing}}

@article{louis2016bias,title={Bias-free spin-wave phase shifter for magnonic logic},author={Louis, Steven and Lisenkov, Ivan and Nikitov, Sergei and Tyberkevych, Vasyl and Slavin, Andrei},journal={AIP Advances},volume={6},number={6},year={2016},publisher={AIP Publishing}}

@article{lake2022microscopic,title={Microscopic nonlinear magnonic phase shifters based on ultrathin films of a magnetic insulator},author={Lake, SR and Divinskiy, B and Schmidt, G and Demokritov, SO and Demidov, VE},journal={Applied Physics Letters},volume={121},number={5},year={2022},publisher={AIP Publishing}}

@article{ustinov2021induced,title={Induced nonlinear phase shift of spin waves for magnonic logic circuits},author={Ustinov, Alexey B and Kuznetsov, Nikolai A and Haponchyk, Roman V and L{\"a}hderanta, Erkki and Goto, Taichi and Inoue, Mitsuteru},journal={Applied Physics Letters},volume={119},number={19},year={2021},publisher={AIP Publishing}}

@article{au2012nanoscale,title={Nanoscale spin wave valve and phase shifter},author={Au, Y and Dvornik, Mykola and Dmytriiev, O and Kruglyak, VV},journal={Applied Physics Letters},volume={100},number={17},year={2012},publisher={AIP Publishing}}

@article{zhu2014magnonic,title={Magnonic crystals-based tunable microwave phase shifters},author={Zhu, Y and Chi, KH and Tsai, CS},journal={Applied Physics Letters},volume={105},number={2},year={2014},publisher={AIP Publishing}}

@article{mortada2025nonreciprocal,title={Nonreciprocal Spin Waves in Out-of-Plane Magnetized Coupled Waveguides Reconfigured by Domain Wall Displacements},author={Mortada, Hanadi and Verba, Roman and Wang, Qi and Pirro, Philipp and Hamadeh, Alexandre Abbass},journal={Advanced Electronic Materials},pages={e00575},year={2025},publisher={Wiley Online Library}}

@article{wang2018electric,title={Electric field controlled spin waveguide phase shifter in YIG},author={Wang, Xi-guang and Chotorlishvili, L and Guo, Guang-hua and Berakdar, J},journal={Journal of Applied Physics},volume={124},number={7},year={2018},publisher={AIP Publishing}}

@article{dobrovolskiy2019spin,title={Spin-wave phase inverter upon a single nanodefect},author={Dobrovolskiy, Oleksandr V and Sachser, Roland and Bunyaev, Sergey A and Navas, David and Bevz, Volodymyr M and Zelent, Mateusz and Śmigaj, Wojciech and Rych{\l}y, Justyna and Krawczyk, Maciej and Vovk, Ruslan V and others},journal={ACS applied materials \& interfaces},volume={11},number={19},pages={17654--17662},year={2019},publisher={ACS Publications}}

@article{toniato2022magnonic,title={Magnonic phase shifters for spin based logic devices},author={Toniato, Alberto},journal={Politecnico Milano 1863},year={2022}}

@article{petrillo2024micromagnetic,title={Micromagnetic simulations for local phase control of propagating spin waves through voltage-controlled magnetic anisotropy},author={Petrillo, Adrien and Fattouhi, Mouad and Di Pietro, Adriano and Alerany Sol{\'e}, Marta and Lopez-Diaz, Luis and Durin, Gianfranco and Koopmans, Bert and Lavrijsen, Reinoud and others},journal={Applied Physics Letters},volume={124},number={19},year={2024},publisher={AIP Publishing}}

@article{wojewoda2023phase,title={Phase-resolved optical characterization of nanoscale spin waves},author={Wojewoda, Ond{\v{r}}ej and Hrto{\v{n}}, Martin and Dhankhar, Meena and Kr{\v{c}}ma, Jakub and Dav{\'\i}dkov{\'a}, Krist{\`y}na and Kl{\'\i}ma, Jan and Holobr{\'a}dek, Jakub and Ligmajer, Filip and {\v{S}}ikola, Tom{\'a}{\v{s}} and Urb{\'a}nek, Michal},journal={Applied Physics Letters},volume={122},number={20},year={2023},publisher={AIP Publishing}}

@article{rousseau2015realization,title={Realization of a micrometre-scale spin-wave interferometer},author={Rousseau, O and Rana, B and Anami, R and Yamada, M and Miura, K and Ogawa, S and Otani, Y},journal={Scientific reports},volume={5},number={1},pages={9873},year={2015},publisher={Nature Publishing Group UK London}}

@article{demidov2009control,title={Control of spin-wave phase and wavelength by electric current on the microscopic scale},author={Demidov, Vladislav E and Urazhdin, Sergei and Demokritov, Sergej O},journal={Applied Physics Letters},volume={95},number={26},year={2009},publisher={AIP Publishing}}

@article{kostylev2007resonant,title={Resonant and nonresonant scattering of dipole-dominated spin waves from a region of inhomogeneous magnetic field in a ferromagnetic film},author={Kostylev, MP and Serga, AA and Schneider, T and Neumann, T and Leven, B and Hillebrands, B and Stamps, RL},journal={Physical Review B—Condensed Matter and Materials Physics},volume={76},number={18},pages={184419},year={2007},publisher={APS}}

@article{han2019mutual,title={Mutual control of coherent spin waves and magnetic domain walls in a magnonic device},author={Han, Jiahao and Zhang, Pengxiang and Hou, Justin T and Siddiqui, Saima A and Liu, Luqiao},journal={Science},volume={366},number={6469},pages={1121--1125},year={2019},publisher={American Association for the Advancement of Science}}

@article{qin2021electric,title={Electric-field control of propagating spin waves by ferroelectric domain-wall motion in a multiferroic heterostructure},author={Qin, Huajun and Dreyer, Rouven and Woltersdorf, Georg and Taniyama, Tomoyasu and van Dijken, Sebastiaan},journal={Advanced Materials},volume={33},number={27},pages={2100646},year={2021},publisher={Wiley Online Library}}

@article{zhao2020spin,title={Spin waves and transverse domain walls driven by spin waves: Role of damping},author={Zhao, Zi-Xiang and He, Peng-Bin and Cai, Meng-Qiu and Li, Zai-Dong},journal={Chinese Physics B},volume={29},number={7},pages={077502},year={2020},publisher={IOP Publishing}}

@article{yan2011all,title={All-magnonic spin-transfer torque and domain wall propagation},author={Yan, Peng and Wang, XS and Wang, XR},journal={Physical review letters},volume={107},number={17},pages={177207},year={2011},publisher={APS}}

@article{hertel2004domain,title={Domain-wall induced phase shifts in spin waves},author={Hertel, Riccardo and Wulfhekel, Wulf and Kirschner, J{\"u}rgen},journal={Physical review letters},volume={93},number={25},pages={257202},year={2004},publisher={APS}}

@article{buijnsters2016chirality,title={Chirality-dependent transmission of spin waves through domain walls},author={Buijnsters, FJ and Ferreiros, Yago and Fasolino, A and Katsnelson, Mikhail I},journal={Physical review letters},volume={116},number={14},pages={147204},year={2016},publisher={APS}}

@article{fan2023coherent,title={Coherent magnon-induced domain-wall motion in a magnetic insulator channel},author={Fan, Yabin and Gross, Miela J and Fakhrul, Takian and Finley, Joseph and Hou, Justin T and Ngo, Steven and Liu, Luqiao and Ross, Caroline A},journal={Nature Nanotechnology},volume={18},number={9},pages={1000--1004},year={2023},publisher={Nature Publishing Group UK London}}

@article{chang2018ferromagnetic,title={Ferromagnetic domain walls as spin wave filters and the interplay between domain walls and spin waves},author={Chang, Liang-Juan and Liu, Yen-Fu and Kao, Ming-Yi and Tsai, Li-Zai and Liang, Jun-Zhi and Lee, Shang-Fan},journal={Scientific reports},volume={8},number={1},pages={3910},year={2018},publisher={Nature Publishing Group UK London}}

@article{kim2012interaction,title={Interaction between propagating spin waves and domain walls on a ferromagnetic nanowire},author={Kim, J-S and St{\"a}rk, Martin and Kl{\"a}ui, Mathias and Yoon, J and You, C-Y and Lopez-Diaz, Luis and Martinez, E},journal={Physical Review B—Condensed Matter and Materials Physics},volume={85},number={17},pages={174428},year={2012},publisher={APS}}

@article{vansteenkiste2014design,title={The design and verification of MuMax3},author={Vansteenkiste, Arne and Leliaert, Jonathan and Dvornik, Mykola and Helsen, Mathias and Garcia-Sanchez, Felipe and Van Waeyenberge, Bartel},journal={AIP advances},volume={4},number={10},year={2014},publisher={AIP Publishing}}

@article{nikitin2014all,title={All-thin-film multilayered multiferroic structures with a slot-line for spin-electromagnetic wave devices},author={Nikitin, Andrey A and Ustinov, Alexey B and Semenov, Alexander A and Kalinikos, Boris A and L{\"a}hderanta, E},journal={Applied Physics Letters},volume={104},number={9},year={2014},publisher={AIP Publishing}}

@article{morozova2020magnonic,title={Magnonic crystal-semiconductor heterostructure: Double electric and magnetic fields control of spin waves properties},author={Morozova, MA and Romanenko, DV and Serdobintsev, AA and Matveev, OV and Sharaevskii, Yu P and Nikitov, SA},journal={Journal of Magnetism and Magnetic Materials},volume={514},pages={167202},year={2020},publisher={Elsevier}}

@article{wang2026coherent,title={Coherent Microwave Driving of Domain Wall Depinning in a Ferrimagnetic Garnet},author={Wang, Hanchen and van Schie, Laura and Erickson, Adam and Riddiford, Lauren J and Petrosyan, Davit and Degen, Christian L and Schlitz, Richard and Legrand, William and Gambardella, Pietro},journal={arXiv preprint arXiv:2604.19164},year={2026}}

@article{tetienne2014nanoscale,title={Nanoscale imaging and control of domain-wall hopping with a nitrogen-vacancy center microscope},author={Tetienne, J-P and Hingant, T and Kim, J-V and Diez, L Herrera and Adam, J-P and Garcia, K and Roch, J-F and Rohart, S and Thiaville, A and Ravelosona, D and others},journal={Science},volume={344},number={6190},pages={1366--1369},year={2014},publisher={American Association for the Advancement of Science}}

@article{hayashi2008current,title={Current-controlled magnetic domain-wall nanowire shift register},author={Hayashi, Masamitsu and Thomas, Luc and Moriya, Rai and Rettner, Charles and Parkin, Stuart SP},journal={Science},volume={320},number={5873},pages={209--211},year={2008},publisher={American Association for the Advancement of Science}}

@article{parkin2008magnetic,title={Magnetic domain-wall racetrack memory},author={Parkin, Stuart SP and Hayashi, Masamitsu and Thomas, Luc},journal={science},volume={320},number={5873},pages={190--194},year={2008},publisher={American Association for the Advancement of Science}}

@article{chen202250,title={The 50 nm-thick yttrium iron garnet films with perpendicular magnetic anisotropy},author={Chen, Shuyao and Xie, Yunfei and Yang, Yucong and Gao, Dong and Liu, Donghua and Qin, Lin and Yan, Wei and Tan, Bi and Chen, Qiuli and Gong, Tao and others},journal={Chinese Physics B},volume={31},number={4},pages={048503},year={2022},publisher={Chinese Physical Society and IOP Publishing Ltd}}

@article{lin2020bi,title={Bi-YIG ferrimagnetic insulator nanometer films with large perpendicular magnetic anisotropy and narrow ferromagnetic resonance linewidth},author={Lin, Yaning and Jin, Lichuan and Zhang, Huaiwu and Zhong, Zhiyong and Yang, Qinghui and Rao, Yiheng and Li, Mingming},journal={Journal of Magnetism and Magnetic Materials},volume={496},pages={165886},year={2020},publisher={Elsevier}}
\end{document}